\documentclass[]{ceurart}
\usepackage[T1]{fontenc}
\usepackage{graphicx}
\usepackage{hyperref}
\usepackage{color}

\usepackage{xspace}
\usepackage{mathtools}
\usepackage{todonotes}
\usepackage{multirow,changepage}
\usepackage{hhline}
\usepackage{longtable}
\usepackage{quoting,xparse}
\usepackage{amsmath, amssymb}
\usepackage{fancybox}
\usepackage{array}
\usepackage{multicol}
\usepackage[linesnumbered,ruled,noend]{algorithm2e}
\usepackage[inline]{enumitem}
\usepackage{blkarray}
\usepackage{siunitx}
\usepackage{booktabs}
\usepackage{listings}

\newtheorem{theorem}{Theorem}

\newtheorem{example}[theorem]{Example}

\def\C#1{`\hbox to .8em{\hfill\texttt{#1}\hfill}'} 
\def\P{\hbox to 1em{\hfill$\mid$\hfill}}           

\newcommand{\KK}{\mathbb{K}}

\def\<#1>{\langle#1\rangle}
\def\Var{\operatorname{Var}}

\DeclareMathOperator\ext{ext}

\usepackage{mdframed}

\def\orcid#1{\smash{\href{http://orcid.org/#1}{\protect\raisebox{-1.25pt}{\protect\includegraphics{orcid_color.eps}}}}}

\newcommand\rulename[1]{\textsc{#1}}
\newcommand{\AxiomRuleLbl}{\rulename{Axiom}\xspace}

\newcommand{\DelRuleLbl}{\rulename{Deletion}\xspace}
\newcommand{\LinComRuleLbl}{\rulename{LinComb}\xspace}
\newcommand{\PatternNewRuleLbl}{\rulename{PatternNew}\xspace}
\newcommand{\PatternApplyRuleLbl}{\rulename{PatternApply}\xspace}
\newcommand{\ExtRuleLbl}{\rulename{Ext}\xspace}

\newcommand\pachecktwo{\textsc{Pacheck}~2.0\xspace}

\begin{document}
\copyrightyear{2025}
\copyrightclause{Copyright for this paper by its authors.
  Use permitted under Creative Commons License Attribution 4.0
  International (CC BY 4.0).}

\conference{}
\title{Recycling Algebraic Proof Certificates}
%
%

\author[1]{Daniela Kaufmann}[%
orcid=0000-0002-5645-0292,
email=daniela.kaufmann@tuwien.ac.at,
url=https://danielakaufmann.at/,
]
\cormark[1]
\address[1]{TU Wien, Vienna, Austria}

\author[2]{Clemens Hofstadler}[%
orcid=0000-0002-3025-0604,
email=clemens.hofstadler@jku.at,
url=https://clemenshofstadler.com/,
]
\cormark[1]
\address[2]{Johannes Kepler University, Linz, Austria}
 
\cortext[1]{Corresponding author.}

\begin{abstract}
  Proof certificates can be used to validate the correctness of algebraic derivations. However, in practice, we frequently observed that the exact same proof steps are repeated for different sets of variables, which leads to  unnecessarily large proofs. 
  To overcome this issue we extend the existing  Practical Algebraic Calculus with linear combinations (LPAC)
  with two new proof rules that allow us to capture and
  reuse parts of the proof to derive a more condensed proof certificate. We
  integrate these rules into the proof checker \pachecktwo. Our experimental
  results demonstrate that the proposed extension helps to reduce both
  proof size and verification time.
 
\end{abstract}


\maketitle
\begin{mdframed}
\textbf{Setting:}
\emph{Let $X$ be a set of variables, $\KK$ a field. Let $G \subseteq \KK[X]$, $f \in \KK[X]$.
The goal is to derive a proof certificate validating that $f \in \<G>$.}
\end{mdframed}

Although our proposed idea of proof recycling works in this general setting, our main focus is on applications where all variables $X$ are
Boolean.
Recall that this can be modeled by adding, for each variable $x \in X$, the Boolean value polynomial $x(x-1) = x^{2} - x$ to $G$. 
Therefore, we assume that $B(X) = \{x^2-x \mid x \in X\} \subseteq~G$. 
Our proof rules (and the existing rules of LPAC; see Figure~\ref{fig:lpacrules})
are instantiated for this particular use case, but they can be easily adapted to the general~setting. 

Two proof systems are commonly studied for our application, \emph{polynomial
calculus} (PC)~\cite{CleggImpagliazzo-ACM1996}, and \emph{Nullstellensatz}
(NSS)~\cite{BeameImpagliazzoKrajicekPitassiPudlak-LMS96}. NSS proofs represent~$f$ as a linear combination of the polynomials in $G$. However, if the
expansion of the linear combination is not equal to $f$, it is unclear how to
locate the error in the proof. This limits the usefulness of a NSS proof for
debugging. In contrast, proofs in PC capture whether $f$ can be derived from
$G$ using proof steps from ideal theory. However, PC as originally
defined~\cite{CleggImpagliazzo-ACM1996} is not suitable for effective proof
checking, because information of the origin of the proof steps is missing.

The Practical Algebraic Calculus (PAC)~\cite{RitircBiereKauers-SCSC18}
addressed this by explicitly encoding each polynomial operation, enabling
stepwise verification and hence error localization. In a later work~\cite{KaufmannFleuryBiereKauers-FMSD22}, PC was combined with NSS in the proof calculus LPAC (PAC with linear combinations) to yield, shorter, yet traceable proofs.

However, in some applications, the same proof steps are repeated over and over
again for different variable instances. For example, in \emph{arithmetic
  circuit verification} it is required to reason over structurally equivalent
building blocks, such as full- and half-adders, of the circuit. Until now (e.g., in~\cite{KaufmannBiere-TACAS21}), all
the proof steps for these building blocks were explicitly recomputed from scratch, 
despite differing only in variable names, rather than being cached and reused.

We now propose to extend LPAC with two new proof rules \PatternNewRuleLbl and
\PatternApplyRuleLbl, which allow us to save a pattern in the proof that can be
reinstated with different polynomials to recycle parts of the proof.


\section{LPAC}

\begin{figure}[tb]
  \begin{minipage}{0.95\textwidth}
    \begin{multicols}{2} 

      \noindent
      \textbf{\AxiomRuleLbl($i, p$)}

      \vspace*{-.75\baselineskip}\noindent\rule{\columnwidth}{0.4pt}
    \vspace*{-2.25\baselineskip}

    \begin{center}\large
        $\begin{array}{l}
            \frac{(X, P) \quad P(i) = \bot \quad p \in \KK[X]}{(X \cup \Var(p), P(i \mapsto p))}
          \end{array}$
        \end{center}

      \noindent
   \textbf{\DelRuleLbl($i$)}

      \vspace*{-.75\baselineskip}\noindent\rule{\columnwidth}{0.4pt}
      \vspace*{-.75\baselineskip}

      \centering\large
      $\begin{array}{l}
          \frac{(X, P)}{(X, P(i \mapsto \bot))}
        \end{array}$

    \end{multicols}

    \vspace{3ex}
    \noindent
    \textbf{\LinComRuleLbl($i, (j_1, \ldots, j_n), (q_1, \ldots, q_n), p$)}

    \vspace*{-.75\baselineskip}\noindent\rule{\columnwidth}{0.4pt}
    \vspace*{-1.25\baselineskip}

    \begin{center}
      \large
    $\begin{array}{l}
        \frac{(X, P) \quad P(j_1) \neq \bot, \ldots, P(j_n) \neq \bot, P(i) = \bot \quad
          p, q_1, \ldots, q_n \in \KK[X] \quad
          p = q_1 \cdot P(j_1) + \cdots + q_n \cdot P(j_n) \text{ mod} \langle B(X) \rangle}{(X, P(i \mapsto p))}
      \end{array}$
    \end{center}

    \medskip

     \vspace{2ex}
    \noindent
   \textbf{\ExtRuleLbl($i, v, q$)}

    \vspace*{-.75\baselineskip}\noindent\rule{\textwidth}{0.4pt}
    \vspace*{-.75\baselineskip}

    \centering
    \large
    $\begin{array}{l}
        \frac{(X, P) \quad P(i) = \bot \quad v \notin X \quad q \in \KK[X] \quad q^2 - q \in \langle B(X) \rangle}{(X \cup \{v\}, P(i \mapsto -v + q))}
      \end{array}$

  \end{minipage}

\vspace{2ex}
  \caption{Existing Proof Rules of LPAC}
  \label{fig:lpacrules}
\end{figure}

An LPAC proof certificate can be used to validate that a polynomial $f \in \KK[X]$ can be derived from a given set of polynomials $G \subseteq \KK[X]$ (= the proof axioms) using polynomial addition and multiplication. The semantics of LPAC can be seen in Figure~\ref{fig:lpacrules}. A more detailed explanation including correctness arguments is given in~\cite{KaufmannFleuryBiereKauers-FMSD22}.

Let $X$ denote a set of Boolean variables and let $P$ be a sequence of polynomials, which can be accessed via indices. An LPAC proof is a sequence of states $(X,P)$, with the initial state $(\{\}, [])$. At any state, $P$ will only contain polynomials that are either axioms or can be derived from polynomials already contained in $P$ using the inference rules of the calculus.

Using the \AxiomRuleLbl rule, we add polynomials $p$ to $P$.
The \DelRuleLbl rule allows us to remove polynomials from $P$ when they are no
longer needed, which helps to save memory during checking. 
The \LinComRuleLbl rule  can be used to add polynomials to $P$ that can be derived as a linear combination of polynomials already stored in $P$. 
Using the \ExtRuleLbl rule we can add polynomials to the proof, which can be used to introduce placeholder variables $v$ for Boolean expressions~$q$.

\ \\
\emph{Proof Checking:}
LPAC proofs can be checked using, e.g., \pachecktwo~\cite{KaufmannFleuryBiereKauers-FMSD22}. Proof
checking is applied on the fly. For instance, after parsing a \LinComRuleLbl proof step, \pachecktwo checks whether all input polynomials exist, 
calculates the linear combination of the given input, and compares it to the given conclusion polynomial $p$ of the \LinComRuleLbl rule. 
Additionally, it is checked whether one of the conclusion polynomials of a \LinComRuleLbl step is equal to the target $f$ at any point in the proof.


\section{Recycling Proof Steps}\label{sec:proof}

In the following, we introduce the new proof rules \PatternNewRuleLbl and
\PatternApplyRuleLbl to add and instantiate proof patterns.
In principle, these rules could be generalized and used to extend any polynomial proof system, including those that do not work over Boolean variables.
Here, we focus on extending LPAC and instantiate the new rules to work for this particular proof system. 
For that, we also extend an LPAC proof from $(X, P)$ to $(X,P,C)$, where $C$ denotes a sequence of patterns. 
Initially $C~=~[]$. Note that $C$ cannot be changed by any of the rules in Figure~\ref{fig:lpacrules}.

\ \\
\emph{\PatternNewRuleLbl: }
A pattern is identified with a unique index $i$ that will be used in the \PatternApplyRuleLbl rule to identify the pattern. A pattern has three components: \emph{sequence of
  inputs $I$}, \emph{sequence of proof steps $S$}, and \emph{set of outputs~$O$}. 
 
The sequence of inputs represents the axioms of the pattern. The proof
steps $S$ form a self-contained proof fragment. In
``Correct-LPAC-Proof($I,S$)'' it is checked whether the proof steps in $S$ can be
derived by the LPAC inference rules using the axioms given in $I$. The set of outputs $O$ contains those
conclusion polynomials of the proof steps in $S$, which will be stored as
output polynomials of the pattern. 

We explicitly extract any extension variables $V_{\ext}$ from the outputs $O$ of the pattern that have been added in the proof steps $S$. When applying the pattern, these variables have to be instantiated as fresh variables. On the other hand, it is not necessary to store any internal  extension variables of $S$ that are not contained in any polynomial in $O$, since they are only used internally to derive the outputs $O$. They are not needed for the application of the pattern.

After a pattern is checked for correctness, we only store $I$, $O$, and $V_{\ext}$ in $C$. The proof steps $S$ are discarded.
On a practical note,  since a pattern can be seen as a standalone LPAC
proof, we are free to (re-)use any variables and indices.

\begin{figure}[tb]
  \begin{minipage}{0.95\textwidth}

    \noindent
    \textbf{\PatternNewRuleLbl($i,I, S, O$)}

    \vspace*{-.75\baselineskip}\noindent\rule{\columnwidth}{0.4pt}
    \vspace*{-1.\baselineskip}

    \begin{center}\large
    $\begin{array}{l}
        \frac{(X, P, C) \quad C(i) = \bot \quad \text{Correct-LPAC-Proof}(I, S) \quad O \subseteq \text{Conclusions}(S) \quad
        V_{\ext}= \text{ExtensionVar}(O)}{(X, P, C \cup (i, I, O,V_{\ext}))}
      \end{array}$
    \end{center}

    \medskip

     \vspace{2ex}
    \noindent
    \textbf{\PatternApplyRuleLbl($i, X_{\ext}, \varphi,  (j_1, \ldots, j_m), \{(k_1, p_1), \ldots, (k_n, p_n)\}$)}

    \vspace*{-.75\baselineskip}\noindent\rule{\textwidth}{0.4pt}
    \vspace*{-.25\baselineskip}

    \centering
    $\begin{array}{l}
        \frac{\scriptsize\begin{array}{c}
                       (X, P, C) \qquad 
                       C(i)\neq \bot \qquad 
                        X_{\ext} \not\subseteq X \qquad
                       \forall v \in C(i).V_{\ext}: \varphi(v) \in X_{\ext} \quad
                       \varphi \text{ injective on } C(i).V_{\ext} \\[0.2em]
                       \forall v \in \Var(C(i)): \varphi(v)^{2} - \varphi(v) \in \<B(X\cup X_{\ext})> \qquad
                       P(j_1) \neq \bot, \ldots, P(j_m) \neq \bot \quad \quad
                       C(i).I =_\varphi \{P(j_1),\ldots, P(j_m)\} \\[0.2em]
                       P(k_1) = \bot, \ldots, P(k_n) = \bot  \quad \quad
                       p_1,\ldots,p_n \in \KK[X \cup X_{\ext}] \quad \quad
                       C(i).O =_\varphi \{p_1, \ldots, p_n\} 
                      \end{array}}{(X \cup X_{\ext}, P(k_1 \mapsto p_1, \ldots, k_n \mapsto p_n), C)}
      \end{array}$
  \end{minipage}
  \vspace{2ex}
  \caption{New Proof Rules of LPAC}
  \label{fig:newlpacrules}
\end{figure}

\ \\
\emph{\PatternApplyRuleLbl: }
The index $i$ refers to a pattern in $C$.
The set $X_{\ext}$ contains all extension variables that will be added to the LPAC proof through the application of the pattern. These variables are not allowed to be contained in the set of variables $X$ of the current proof state $(X,P,C)$.

The mapping $\varphi$ connects the pattern $C(i)$ to the LPAC proof $P$ by mapping all variables (including those in $V_{\ext}$) in $C(i)$ 
to polynomials over the variables $X \cup X_{\ext}$.
This mapping has to satisfy two conditions:
First, it has to injectively map the extension variables used in $C(i)$ to the new extension variables in $X_{\ext}$, that is,
for all $v_{1},v_{2} \in V_{\ext}$, we must have $\varphi(v_{1}), \varphi(v_{2})  \in X_{\ext}$ and $\varphi(v_{1}) \neq \varphi(v_{2})$ if $v_{1} \neq v_{2}$. 
Moreover, $\varphi$ has to comply with the Boolean axioms in the following way:
for each variable $v$ in $C(i)$, its image under $\varphi$ must satisfy the condition $\varphi(v)^{2} - \varphi(v) \in \langle B(X \cup X_{\ext})\rangle$.
This restriction comes from the fact that all variables are assumed Boolean in an LPAC proof and 
the Boolean axioms are used implicitly (see, for instance, the \LinComRuleLbl rule in Figure~\ref{fig:lpacrules}).
In general, when a proof system treats certain relations implicitly, the mapping $\varphi$ has to be consistent with those relations.
In contrast, for a proof system that does not use any implicit relations, $\varphi$ can map the variables from $C(i)$ to arbitrary polynomials.

The tuple $(j_1, \ldots, j_m)$ contains the indices to those polynomials of $P$ that are used as inputs to the pattern. 
We verify that these polynomials match the pattern’s inputs under the mapping $\varphi$, denoted by $=_\varphi$ in Figure~\ref{fig:newlpacrules}.
We do an analogous check for all the output polynomials $p_i$ in the set $\{(k_1, p_1), \ldots, (k_n, p_n)\}$. 
If the pattern is correctly applied, we store the polynomials $p_i$ as correctly derived polynomials~in~$P$, i.e., we set $P(k_i) \mapsto p_i$.

\ \\
\emph{Proof Checking: }
For the \PatternNewRuleLbl rule, we have to apply the following checks during proof checking: 
\begin{enumerate}[label=(\roman*),topsep=10pt,itemsep=-1ex,partopsep=3ex,parsep=1ex]
\item do we already have a pattern with index $i$;
\item do $I$ and $S$ form a correct standalone LPAC proof; 
\item are all polynomials in $O$ conclusion polynomials of proof steps in $I$ or $S$. 
\end{enumerate}

\newpage
For the \PatternApplyRuleLbl rule, we have to apply the following checks:
\begin{enumerate}[label=(\roman*),topsep=10pt,itemsep=-1ex,partopsep=3ex,parsep=1ex]
\item does pattern $C(i)$ exist;
\item  are all variables in $X_{\ext}$ fresh variables that are not contained in $X$;
\item does the mapping $\varphi$ map the extension variables of the pattern injectively to the fresh variables in~$X_{\ext}$;
\item does the mapping $\varphi$ comply with the Boolean axioms $B(X \cup X_{\ext})$; 
\item do the input polynomials $I$ correspond to the input polynomials of $C(i)$ after the mapping $\varphi$ is applied;
\item do the output polynomials~$O$ correspond to the output polynomials of $C(i)$ after $\varphi$ is applied.
\end{enumerate}

\section{Examples}

We illustrate the new proof rules on two examples. 
In the first example we show how patterns can help to reduce the size of the proof. 
In the second example we show how to treat extension variables in proof patterns.

\begin{example}\label{ex:lpac_extended}
  Let $G=\{ x+2y-2, -y-z+1, a + 2b-2, -b-z+1, a-x+1\}$. 
  We use the \PatternNewRuleLbl and \PatternApplyRuleLbl rules to derive $ 1 \in \<G>$ in an LPAC proof. 
  We use $p_i$ as indices for the proof steps inside the pattern and $l_i$ as indices for the proof steps outside the pattern.
  

    {\small\center
        \begin{lstlisting}[mathescape]      
1 $\AxiomRuleLbl$($l_1$, $x+2y-2$)                       
2 $\AxiomRuleLbl$($l_2$, $-y-z+1$)  
3 $\AxiomRuleLbl$($l_3$, $a + 2b-2$)
4 $\AxiomRuleLbl$($l_4$, $-b-z+1$)
5 $\AxiomRuleLbl$($l_5$, $a-x+1$)                         
6 $\PatternNewRuleLbl$($1$, [($p_1$ $v_1-2v_2$), ($p_2$ $v_2-v_3$)], 
          [$\LinComRuleLbl$($p_3$, ($p_1,p_2$), ($1$,$2$), $v_1 - 2v_3$)], {$p_3$})
7 $\PatternApplyRuleLbl$($1$, {}, [$v_1 \mapsto x$, $v_2 \mapsto 1-y$, $v_3 \mapsto z$], ($l_1,l_2$), {($l_6$, $x - 2z$)})         
8 $\PatternApplyRuleLbl$($1$, {}, [$v_1 \mapsto a$, $v_2 \mapsto 1-b$, $v_3 \mapsto z$], ($l_3,l_4$), {($l_7$, $a - 2z$)})
9 $\LinComRuleLbl$($l_8$, ($l_{5},l_6,l_7$), ($1$,$1$,$-1$), $1$)
  \end{lstlisting}
    }

The pattern created in this proof captures the derivation of a specific linear combination of two polynomials:
Starting from $p_{1} = v_{1} - 2v_{2}$ and $p_{2} = v_{2} - v_{3}$, defined over variables $v_{1}, v_{2}, v_{3}$,
we can derive the linear combination $p_{3} = 
p_{1} + 2p_{2} = v_{1} - 2v_{3}$.
We then apply this pattern in two different ways.

First, we instantiate $v_{1}, v_{2}, v_{3}$ by $x$, $1-y$, and $z$, respectively.
Note that this substitution complies with the Boolean axioms.  
For instance, $\varphi(v_{2})^{2} - \varphi(v_{2}) = (1-y)^{2} - (1-y) = y^{2} - y \in \langle B(\{a,b,x,y,z\})\rangle$.
Under this substitution, $p_{1}$ and $p_{2}$ reduce to the axioms $l_{1}$ and $l_{2}$, and the output polynomial $p_{3}$ 
becomes $x - 2z$.
In the second application, we substitute $v_{1}, v_{2}, v_{3}$ with $a$, $1-b$, and $z$, respectively.
Then $p_{1}$ and $p_{2}$ reduce to the axioms $l_{3}$ and $l_{4}$, and the output polynomial $p_{3}$ 
becomes $a - 2z$.

By using the pattern, we have to compute the linear combination $p_{1} + 2p_{2}$ only once and can then reuse it through instantiation in each application.
In contrast, a classical LPAC proof would require computing each linear combination $l_{1} + 2l_{2}$ and $l_{3} + 2l_{4}$ separately. 
\end{example}

\begin{example}\label{ex:lpac_resolution}
  In this example, we algebraically mimic the Boolean resolution rule. That is, from 
  $\neg x \lor \neg y$ and $y \lor z$, we derive $\neg x \lor z$.
 Algebraically, the two axioms can be encoded as $G=\{ xy, yz-y-z+1\}$. The goal is to derive $x\bar{z} \in \<G \cup \{\bar{z} - (1-z)\}>$, with $\bar{z}$ being a new variable representing $\bar{z} = 1-z$.

    {\small\center
        \begin{lstlisting}[mathescape]      
1 $\AxiomRuleLbl$($l_1$, $xy$)                       
2 $\AxiomRuleLbl$($l_2$, $yz-y-z+1$)  
3 $\PatternNewRuleLbl$($1$, [($p_1$ $v_1v_2$), ($p_2$ $v_2v_3-v_2-v_3+1$)], 
          [$\ExtRuleLbl$($p_3$, ($w_3$), ($1-v_3$)), 
           $\LinComRuleLbl$($p_4$, ($p_1,p_2,p_3$), ($w_3$,$v_1$,$v_1v_2-v_1$), $v_1w_3$)], {$p_3, p_4$})
4 $\PatternApplyRuleLbl$($1$, {$\bar{z}$}, [$v_1 \mapsto x$, $v_2 \mapsto y$, $v_3 \mapsto z$, $w_3 \mapsto \bar{z}$], ($l_1,l_2$), 
           {($l_3$, $-\bar{z}+1-z$), ($l_4$, $x\bar{z}$)})         
  \end{lstlisting}
    }

\end{example}

\section{Experimental Evaluation}

We have extended \pachecktwo with the \PatternNewRuleLbl and \PatternApplyRuleLbl rules and show the effect of adding those rules on a small set of benchmarks that originate from multiplier verification. In that setting, we prove that the circuit specification is contained in the ideal generated by the circuit encoding. 

\emph{Proof Generation:} 
In our recent work on circuit verification~\cite{HofstadlerKaufmann-CP25}, we extract linear relations from subcircuits, which are then used for the ideal membership test. These subcircuits are defined syntactically. Since circuits typically consist of repeated usage of the same building blocks, such as full- and half-adders, we often identify the same subcircuits multiple times. To avoid redundant computations, we started caching these subcircuits.

To detect isomorphic subcircuits, we map all gate variables of a subcircuit to a standardized set of variables and check for syntactic equivalence. Using this approach, we are able to cache over 80\% of the identified subcircuits. For each computation within a subcircuit, we store a corresponding proof pattern. Whenever a cached pattern is retrieved, we instantiate and apply the corresponding proof pattern to avoid the need for re-computation.

In the following, we use a subset of the benchmarks from~\cite{HofstadlerKaufmann-CP25} to demonstrate the effect of using proof patterns. All benchmarks are drawn from the experimental evaluations in~\cite{Kaufmann-TACAS25,HofstadlerKaufmann-CP25}, and are available from the artifact~\cite{kaufmann_2025_14610365} of the paper~\cite{Kaufmann-TACAS25}.

\setlength{\tabcolsep}{5pt}
\begin{table}[tb]
  \centering 
  \begin{tabular}{l|r||r|r|r|r||r|r|r|r|r|r|r}
  \toprule
  & & \multicolumn{4}{c||}{LPAC without Patterns} & \multicolumn{7}{c}{LPAC with Patterns} \\
  Name & Axioms & Steps & File & Mem & Time &  Steps & \# & Apply & max $\lvert S\vert$ & File & Mem & Time \\
    & ($10^3$) & ($10^3$) & (MB) & (MB) & (s) &  ($10^3$) &  &  & & (MB) & (MB) & (s) \\
  \midrule
  abc & 64 & 196 & 454 & 982 & 10.17 & 48 &   9 & 4032 &46 & 439 & 913 & 9.61 \\
abc-rsn & 64 & 196 & 454 & 982 & 10.35 & 48 &  12 & 4033 & 65&  439 & 913 & 9.94 \\
abc-cmp & 64 & 196 & 454 & 982 & 10.45 & 48 &  12 & 4033 & 46 & 439 & 913 & 9.90 \\
sp-ar-rc & 96 & 270 & 676 & 1377 & 15.86 & 107 &  18 & 6509 & 83 & 663 & 1315 & 14.87 \\
sp-bd-rc & 98 & 284 & 1130 & 2098 & 32.33 & 111 &  42 & 6520 & 87& 1117 & 2038 & 29.45 \\
sp-dt-rc & 96 & 292 & 1789 & 3112 & 56.63 & 108 &  58 & 7082 & 84& 1776 & 3056 & 55.75 \\
sp-os-rc & 99 & 289 & 1314 & 2380 & 38.62 & 112 &  52 & 6542 & 87& 1300 & 2321 & 38.48 \\
sp-ar-cl & 108 & 2043 & 959 & 1867 & 25.98 & 1820 &  79 & 3999 & 421& 924 & 1749 & 24.37 \\
\bottomrule
  \end{tabular}
  \caption{Benchmark Results}
  \label{tbl:exp}
  \vspace{-2ex}
  \end{table}

  Table~\ref{tbl:exp}  summarizes our results in two blocks: one for the original version of LPAC (``LPAC without Patterns''), the second one with our two new proof rules (``LPAC with Patterns''). None of the proofs contains \ExtRuleLbl or \DelRuleLbl rules.  
  The second column lists the number of axioms, which is equal for both approaches.  
  The next four columns represent LPAC proofs without patterns. We list the number of linear combination proof steps (``Steps''), the size of the proof file in MB (``File''), the used memory in MB (``Mem''), and the time needed to check the proof (``Time'') in seconds. 
  The next seven columns show the results when our two new proof rules are added. We list the number of \PatternNewRuleLbl steps (``\#''),  the total number of \PatternApplyRuleLbl steps (``Apply''), and the maximal number of steps in a pattern (``max $\lvert S\vert$''). 

It can be seen that we can significantly reduce the number of proof steps when using patterns and we always improve on the file size of the proof and the memory and time used to check the proof. 

Further improving the efficiency of proof checking is part of future work.
In particular, we want to investigate the effects of recently developed techniques for finding short proofs of ideal membership~\cite{HV24}.

\begin{acknowledgments}
  This research was supported by Austrian Science Fund (FWF)
  [10.55776/ESP666] and by the LIT AI Lab funded by the state of Upper Austria.
\end{acknowledgments}
%
%
%
\bibliography{paper}
\end{document}